\documentclass[apj]{emulateapj}
\usepackage{apjfonts}

\slugcomment{ApJ, in press; astro-ph/0402247}
\shorttitle{X-RAY MOON AND CHARGE TRANSFER}
\shortauthors{WARGELIN ET AL.}

\newcommand{\chandra}{{\it Chandra}}
\newcommand{\rosat}{{\it ROSAT\,}}

\newcommand{\ace}{{\it ACE}}
\newcommand{\imp}{{\it IMP-8}}

 \newcommand{\icka}{\ion{C}{5} K$\alpha$}
\newcommand{\iclya}{\ion{C}{6} Ly$\alpha$}
\newcommand{\iclyb}{\ion{C}{6} Ly$\beta$}
\newcommand{\iclyg}{\ion{C}{6} Ly$\gamma$}
 \newcommand{\inka}{\ion{N}{6} K$\alpha$}

 \newcommand{\ioka}{\ion{O}{7} K$\alpha$}
 \newcommand{\iokb}{\ion{O}{7} K$\beta$}
\newcommand{\iolya}{\ion{O}{8} Ly$\alpha$}
\newcommand{\iolyb}{\ion{O}{8} Ly$\beta$}
\newcommand{\iolyg}{\ion{O}{8} Ly$\gamma$}
\newcommand{\iolyd}{\ion{O}{8} Ly$\delta$}
\newcommand{\iolye}{\ion{O}{8} Ly$\epsilon$}
\newcommand{\ineka}{\ion{Ne}{9} K$\alpha$}
\newcommand{\imgka}{\ion{Mg}{11} K$\alpha$}

\begin{document}
\title{{\it CHANDRA}\/ OBSERVATIONS OF THE ``DARK'' MOON AND
GEOCORONAL SOLAR-WIND CHARGE TRANSFER}

\author{B.\ J.\ Wargelin\altaffilmark{1},
M.\ Markevitch\altaffilmark{2},
M.\ Juda,
V.\ Kharchenko,
R.\ Edgar,
and A.\ Dalgarno}
\affil{Harvard-Smithsonian Center for Astrophysics,
60 Garden Street, Cambridge, MA 02138}
\altaffiltext{1}{bwargelin@cfa.harvard.edu}
\altaffiltext{2}{Also IKI, Moscow, Russia}


\begin{abstract}

We have analyzed data from two sets of calibration observations of the Moon
made by the {\it Chandra X-Ray Observatory}.
In addition to obtaining a spectrum of the bright side that
shows several distinct fluorescence lines, we also clearly detect 
time-variable soft X-ray emission, 
primarily \ioka\ and \iolya, when viewing the optically dark side.  
The apparent dark-side brightness varied by at least an order of magnitude,
up to $\sim 2 \times 10^{-6}$ phot s$^{-1}$ arcmin$^{-2}$ cm$^{-2}$
between 500 and 900 eV,
which is comparable to the typical 3/4-keV-band background emission measured
in the \rosat\ All-Sky Survey.
The spectrum is also very similar to background spectra recorded
by \chandra\ in low or moderate-brightness regions of the sky.
Over a decade ago, \rosat\ also detected soft X-rays from the dark side
of the Moon, which were tentatively ascribed
to continuum emission from energetic solar wind electrons 
impacting the lunar surface.
The \chandra\ observations, however, with their better
spectral resolution, combined with contemporaneous measurements
of solar-wind parameters, strongly favor charge transfer 
between highly charged solar-wind ions and 
neutral hydrogen in the Earth's geocorona as the mechanism
for this emission.  We present a theoretical model of
geocoronal emission and show that predicted spectra and 
intensities match the \chandra\ observations very well.
We also model the closely related process of heliospheric charge 
transfer and estimate that the total charge transfer flux observed
from Earth amounts to a significant fraction of the 
soft X-ray background,
particularly in the \rosat\ 3/4-keV band.

\end{abstract}




\section{INTRODUCTION}
\label{sec:intro}

As reported by \citet{cit:schmitt1991},
an image of the Moon in soft X-rays (0.1--2 keV) was
obtained by the {\it R\"{o}ntgen Satellite} (\rosat) using its
Position-Sensitive Proportional Counter (PSPC)
on 1990 June 29.
This striking image
showed an X-ray-bright sunlit half-circle on one side, 
and a much dimmer but not completely dark side outlined by a
brighter
surrounding diffuse X-ray background.  Several origins for
the dark-side emission were considered, but the
authors' favored explanation was 
continuum emission arising from solar wind
electrons sweeping around to the unlit side
and impacting on the lunar surface,
producing thick-target
bremsstrahlung.  
Given the very limited energy resolution of the PSPC,
however, emission from multiple lines could not be ruled out.

A significant problem with the bremsstrahlung model
was explaining how electrons from the general
direction of the
Sun could produce 
events on the opposite side of the Moon, with
a spatial distribution
which was ``consistent with the telescope-vignetted
signal of a constant extended source.''
An elegant alternative explanation would be a
source of X-ray emission
{\em between} the Earth and the Moon, but at the time,
no such source could be envisioned.
If this source were also
time-variable, it would account for the 
Long Term Enhancements (LTEs) seen by \rosat.
These occasional increases in the counting rate of 
the PSPC
are vignetted in the same way 
as sky-background X-rays, indicating an external origin
\citep{cit:snowden1995}.
LTEs are distinct from the particle-induced background,
and are uncorrelated with the spacecraft's
orientation or position (geomagnetic latitude, etc.),
although \citet{cit:freyberg1994} noted
that LTEs appeared to be related, by a then unknown mechanism, to
geomagnetic storms and solar wind variations.
The final \rosat\ All-Sky Survey (RASS) diffuse background
maps
\citep{cit:snowden1995,cit:snowden1997}
removed the LTEs, so
far as possible, by comparing multiple observations of
the same part of the sky, but any constant or 
slowly varying ($\tau \ga 1$ week)
emission arising from whatever was causing the LTEs would remain.

A conceptual breakthrough came with
the \rosat\ observation of comet Hyakutake \citep{cit:lisse1996}
and the suggestion by \citet{cit:cravens1997} that charge transfer (CT)
between the solar wind and neutral gas from the comet
gave rise to the observed X-ray emission.
In solar-wind charge transfer,
a highly charged ion
in the wind (usually oxygen or carbon)
collides with neutral gas (mostly water vapor in the case of comets)
and an electron is transferred from the neutral species
into an excited energy level of the wind ion, 
which then decays and emits an X-ray.
This hypothesis has been proven by subsequent observations of
comets such as C/LINEAR 1999 S4 by \chandra\ \citep{cit:lisse2001}
and Hyakutake by {\it EUVE} \citep {cit:krasno2001}
(see also the review by \citet{cit:cravens2002}),
and is supported by
increasingly detailed spectral models
\citep{cit:khar2003,cit:khar2000}.
A more extensive history of the evolution of the 
solar-wind CT concept can be found in
\citet{cit:cravens2001}
and \citet{cit:robertson2003}.

Citing the cometary emission model, 
\citet{cit:cox1998} 
pointed out that CT must occur throughout the heliosphere
as the solar wind interacts with atomic H and He within
the solar system.
\citet{cit:freyberg1998} likewise presented 
\rosat\ High-Resolution Imager data that provided some evidence for a
correlation between
increases in the apparent intensity of comet Hyakutake
and in the detector background; he further suggested
that this could be caused by charge transfer of the
solar wind with the Earth's atmosphere.
A rough broad-band quantitative analysis by 
\citet{cit:cravens2000} 
predicted that heliospheric emission, along with
CT between the solar wind and neutral H in the Earth's
tenuous outer atmosphere (geocorona), accounts
for up to half
of the observed soft X-ray background (SXRB).
Intriguingly,
results from
the Wisconsin Soft X-Ray Background sky survey \citep{cit:mccammon1990} and 
RASS observations \citep{cit:snowden1995} indicate that
roughly half of the 1/4-keV background comes from a ``local hot plasma.''
\citet{cit:cravens2001} also modeled how variations in solar-wind
density and speed
should affect heliospheric and geocoronal CT emission observed at Earth,
and found strong correlations between the measured solar-wind proton flux
and temporal variations in the \rosat\ counting rate.

In this paper we present definitive spectral evidence for geocoronal 
CT X-ray emission,
obtained in \chandra\ observations of the Moon.
Data analysis is discussed in \S\ref{sec:data}, and 
results are presented in \S\ref{sec:results}.
As we show in \S\ref{sec:interp}, model predictions of
geocoronal CT agree very well with the observed 
\chandra\ spectra.  
In \S\ref{sec:sxrb} we estimate the level of heliospheric CT emission, 
discuss the overall contribution of CT emission to the SXRB, 
and assess the observational prospects for improving
our understanding of this subtle but ubiquitous souce of X-rays.


\section{THE DATA}
\label{sec:data}

\renewcommand{\arraystretch}{1.15}
\begin{deluxetable}{ccccc}
\tablecaption{Observation Information \label{table:obsids}}
\tablehead{
                &                &                & \colhead{Exposure}  & \colhead{{\it Chandra}}       \\
\colhead{ObsID} & \colhead{Date} & \colhead{CCDs} & \colhead{(s)}       & \colhead{Time}
}
\startdata
2469    & 2001 Jul 26   & I23, S23      & 2930  & 112500070--112503000 \\
2487    & 2001 Jul 26   & I23, S23      & 2982  & 112503320--112506302 \\
2488    & 2001 Jul 26   & I23, S23      & 2747  & 112507858--112510605 \\
2489    & 2001 Jul 26   & I23, S23      & 2998  & 112510900--112513898 \\
2490    & 2001 Jul 26   & I23, S23      & 2830  & 112515450--112518280 \\
2493    & 2001 Jul 26   & I23, S23      & 2993  & 112518500--112521493 \\
\\
2468    & 2001 Sep 22   & I23, S123     & 3157  & 117529483--117532640 \\
3368    & 2001 Sep 22   & I23, S123     & 2223  & 117532850--117535073 \\
3370    & 2001 Sep 22   & I23, S123     & 4772  & 117536678--117541450 \\
3371    & 2001 Sep 22   & I23, S123     & 3998  & 117541880--117545878
\enddata
\tablecomments{
\chandra\ time 0 corresponds to the beginning of 1998.
Observations run from July 26 02:01:10--07:58:13 UT,
and September 22 07:04:43--11:37:58 UT.
}
\end{deluxetable}

The Moon was observed with the \chandra\
Advanced CCD for Imaging Spectroscopy
(ACIS) in two series of calibration observations on 2001 July 26 and
September 22 totaling 17.5 and 14 ksec, respectively  
(see Table~\ref{table:obsids}).
The intention was to determine the
intrinsic ACIS detector background by using the Moon to block
all cosmic X-ray emission.  
Four of the ACIS CCDs were used in July (I2 and I3 from the ACIS-Imaging
array, and S2 and S3 from the ACIS-Spectroscopy array),
and the S1 chip was added in September.
Two of the chips, S1 and S3, are back-illuminated (BI)
and have better quantum efficiency at low energies than
the front-illuminated (FI) chips, I2, I3, and S2.
As can be seen in Fig.~\ref{fig:moons}, however,
the BI chips have higher intrinsic background than
the FI chips, and also poorer energy resolution.
Telemetry limits prevented the operation of more CCDs
when using ACIS Very Faint mode, which was desired because
of its particle-background rejection utility
\citep{cit:vikhlininmemo}.

The ACIS detector background was also calibrated in an alternative manner
using Event Histogram Mode (EHM; \citet{cit:billermemo}).
The July Moon and EHM spectra from the S3 ship 
were compared by \citet{cit:maxim2002}
and showed good agreement, although there was a noticeable but statistically
marginal excess
near 600 eV in the Moon data.

The dark-Moon vs EHM comparison strongly supports the assumption
that the high-energy particle 
background inside the detector housing where EHM data are collected
is the same as in the focal position.  With that in mind,
new calibration measurements were made on 2002 September 3
with ACIS operating with its standard imaging setup
in a ``stowed'' position, where it was 
both shielded from the sky and removed from the 
radioactive calibration source in its normal off-duty position.
\citep{cit:maximmemo}.
These data (ObsID 62850, 53 ks)
provide the best available
calibration of the intrinsic detector background
and are used in the analysis that follows.


\subsection{Data Preparation}
\label{sec:dataprep}

All data were
processed to level 1 using \chandra\ Interactive Analysis of
Observations (CIAO) software, Pipeline release 6.3.1, with bad-pixel
filtering.  Start and stop times
for each observation were chosen to exclude spacecraft maneuvers.
Apart from the inclusion of CTI corrections (see below) and
a more aggressive exclusion of any possibly questionable data, 
our data processing
is essentially the same as that described by \citet{cit:maxim2002},
who limited their analysis to the July S3 data.
Here we use data from all chips during both the July and September
observations, and include data from periods of partial dark-Moon
coverage by using spatial filtering (see \S\ref{sec:spatial}).

Although ACIS has thinly aluminized filters to limit optical contamination,
the sunlit side of the Moon was so bright that
an excess bias signal was sometimes produced in the CCDs,
particularly in the I2 and I3 chips that imaged that region
during July.
As described in \citet{cit:maxim2002},
a bias correction to each event's pulse height amplitude was
calculated by averaging the 16 lowest-signal pixels
of the 5$\times$5-pixel Very Faint (VF) mode event island.
ObsID 2469 suffered by far the most optical contamination, so
that all data from the I2 chip during that observation had to be discarded.
The I3 chip also had significant contamination, but it was
small enough to be largely corrected.  
As explained in \S\ref{sec:dataspectra}, however,
I3 data from that observation were also excluded as a precaution.  
Two other July observations
required exclusion of some I2 data because of optical leaks
(930 s in ObsID 2490 and 
1140 s in Obsid 2493), but
in all the remaining data
the typical energy correction
was no more than a few eV,
which is insignificant for our purposes.

To improve the effective energy resolution,
we applied standard charge transfer inefficiency (CTI) corrections,
as implemented in the CIAO tool {\tt acis\_process\_events},
to data from the FI chips.
CTI is much less of a problem in the BI chips,
S1 and S3, and no corrections were made to those chips' data.
Finally, VF-mode filtering was applied to 
all the data \citep{cit:vikhlininmemo}
to reduce the particle-induced detector background.
The ``ACIS stowed'' background data were treated in the same way,
except that no optical-contamination corrections were required.


\subsection{Spatial Extractions}
\label{sec:spatial}


\begin{figure}[b]
\epsscale{0.8}
\plotone{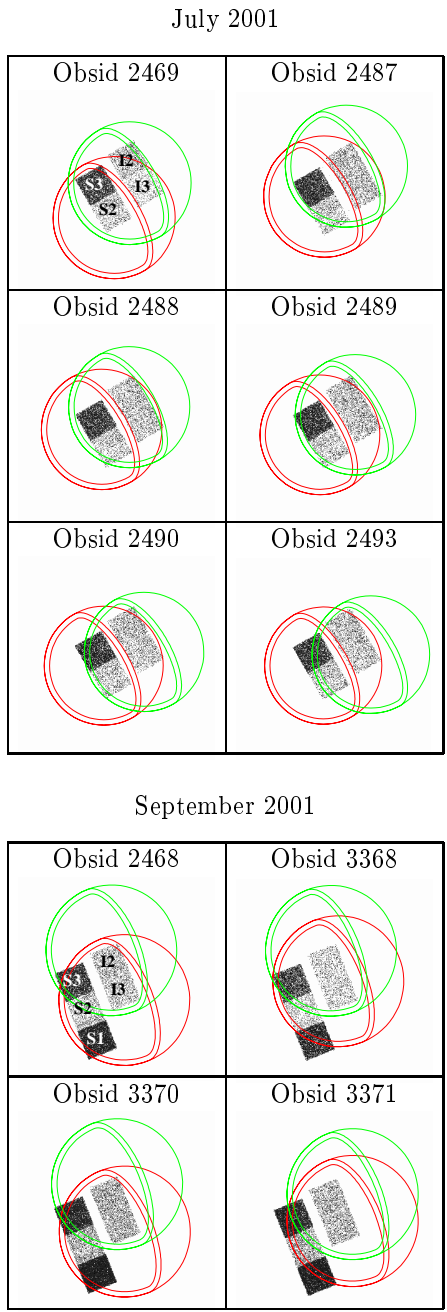}
\vspace{4mm}

\caption{
Moon motion across the ACIS chips during July and September observations.
Green denotes Moon position at
the start of the observation and red at the end.
The illuminated portion of the Moon is the crescent on the right.
Dark-side gibbous extraction regions
provide a 90'' buffer along
the Moon limb and terminator.
S1 and S3 chips are darker because of their higher background.
}
\label{fig:moons}
\end{figure}


As seen in Fig.~\ref{fig:moons}, \chandra\ pointed at a fixed location on the
sky during each observation as the Moon drifted across the field of view.
Using ephemerides for the Moon and {\it Chandra}'s orbit, we 
calculated the apparent position and size of the Moon in
1-minute intervals and extracted data from its dark side,
as well as from the bright side and from the 
unobscured cosmic X-ray background (CXB) within the field of view
for comparison.
Because the Moon moved by up to 16'' per minute,
and to avoid X-ray ``contamination'' of data within each extraction region
(particularly spillover of bright-side or CXB photons into the dark side)
we used generous buffers of
90'' from the terminator and limb for the dark-side extraction,
30'' from the terminator and 60'' from the limb for bright-side data,
and 60'' from the limb for the CXB data.
\chandra\ has a very tight point spread function, with an on-axis
encircled energy fraction of nearly 99\% at 500 eV within 10'';
although off-axis imaging is involved here, 
estimated X-ray contamination is less than $\sim2$\%
within our chosen extraction regions.

Data from all ObsIDs were analyzed in several energy bands
to look for discrete sources but none were found, and lightcurves for
each observation behave as would be expected for uniform emission
within each extraction region.
Effective exposure times (as if one CCD were fully exposed) 
were computed for each ObsID/chip/extraction
combination
by computing extraction areas for each 1-minute interval
(accounting for spacecraft dither, which affects area calculations near 
the chip edges)
and summing the area$\times$time products.
Results are listed in Table~\ref{table:exposures}.

\tabletypesize{\small}
\renewcommand{\arraystretch}{1.1}
\begin{deluxetable*}{cccccccccccc}
\tablecaption{Effective Exposure Times Per Chip\tablenotemark{a} \label{table:exposures}}
\tablehead{
                & \multicolumn{5}{c}{Dark-Side Region}  &       & \multicolumn{5}{c}{Bright-Side Region}\\
\cline{2-6}     \cline{8-12}
\colhead{ObsID} & \colhead{I2} & \colhead{I3} & \colhead{S1} & \colhead{S2} & \colhead{S3} &    & \colhead{I2} & \colhead{I3} & \colhead{S1} & \colhead{S2} & \colhead{S3}
}
\startdata
2469    & 0\tablenotemark{b}    & 0\tablenotemark{c}    &\ldots & 2930  & 2924  &               & 0\tablenotemark{b}    & 0\tablenotemark{c}    &\ldots & 0     & 0     \\
2487    & 1702  & 1759  &\ldots & 2812  & 2982  &               & 722   & 704   &\ldots & 0     & 0     \\
2488    & 1204  & 1233  &\ldots & 2714  & 2747  &               & 979   & 965   &\ldots & 0     & 0     \\
2489    & 1491  & 1375  &\ldots & 2978  & 2977  &               & 984   & 1062  &\ldots & 0     & 0     \\
2490    & 1071\tablenotemark{d} & 1306  &\ldots & 2811  & 2714  &               & 984   & 1062  &\ldots & 0     & 0     \\
2493    & 1456\tablenotemark{e} & 1637  &\ldots & 2862  & 2572  &               & 238   & 847   &\ldots & 0     & 0     \\
\\
Total   & 6924  & 7310  &\ldots & 17107 & 16916 &               & 3503  & 4560  &\ldots & 0     & 0     \\
\tableline\\
2468    & 3157  & 3150  & 1026  & 2534  & 2958  &               & 0     & 0     & 1462  & 177   & 6     \\
3368    & 2223  & 2223  & 162   & 1364  & 2031  &               & 0     & 0     & 1680  & 275   & 0     \\
3370    & 4772  & 4770  & 1281\tablenotemark{f} & 3479  & 3611\tablenotemark{f} &               & 0     & 0     & 2157\tablenotemark{f} & 439   & 20\tablenotemark{f}   \\
3371    & 3995  & 3995  & 265   & 1631  & 1803  &               & 0     & 0     & 3024  & 1114  & 950   \\
\\
Total   & 14147 & 14138 & 2734  & 9008  & 10403 &               & 0     & 0     & 8323  & 2005  & ~~~~~976
\enddata
\tablenotetext{$a$}{~In units of s.~
$^b$~Severe optical leak.  All data excluded.~
$^c$~Unreliable energies and possible event loss from optical leak.  All
data excluded.~ 
$^d$~Optical leak. Time range 112516560--112517490 (930 s) excluded.~
$^e$~Optical leak.  Time range 112520070--112521210 (1140 s) excluded.~
$^f$~Background flare in BI chips.  Time range
  117538500--117538900 (400 s) excluded.
}
\end{deluxetable*}

Because the detector background is not perfectly uniform across each chip,
background data were projected onto the sky
and extracted using the same regions as
for the dark-Moon data.  
Exposure-weighted and epoch-appropriate detector response functions 
(RMFs and ARFs)
were then created using standard CIAO threads,
including the {\tt corrarf} routine, which applies
the ACISABS model to account for contaminant build-up on ACIS.
The detector background rate varies slightly on timescales of months,
so we renormalized the background data to match the corresponding
observational data in the energy range 9.2--12.2 keV, where the
detected signal is entirely from intrinsic background.
The required adjustments were only a few percent.


\subsection{Spectra}
\label{sec:dataspectra}

As described by \citet{cit:maxim2002},
the BI chips, and very rarely the FI chips,
often experience ``soft'' background
flares because of their higher sensitivity to low-energy particles.
A relatively bright flare was found in ObsID 3370, and 400 seconds of
data were removed.
Weaker flares are more common and we judged it better to model and
subtract their small effects rather than exclude large intervals of data.
Soft flares have a consistent spectral shape 
(a powerlaw with high-energy cutoff) and
their intensity at all energies can be determined by integrating the excess
signal (above the ``stowed'' background) in the energy range 2.5--7 keV
where the relative excess is most significant.
We find that spectra from ObsIDs 2468 and 3368 have minor soft-flare
components, and have accounted for them in the results presented later.
An essentially negligible soft-flare excess is also seen and accounted for
when all the July S3 data are combined.


\begin{figure}[b]
\includegraphics[scale=0.47,bb=24 332 530 730,clip]{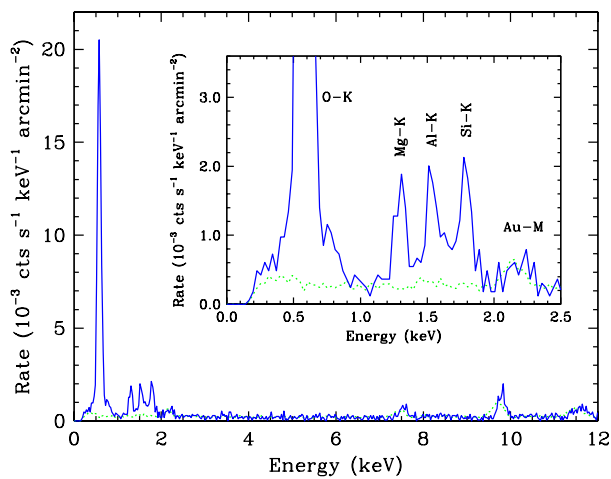}
\caption{
Spectrum of the bright side of the Moon, combining 
I2 and I3 data from all July observations except ObsID 2469,
binned by 2 PI channels (29.2 eV).
Dotted curve is detector background.
Fluorescence K-shell lines from O, Mg, Al, and Si are
shifted up by 50 eV from their true values because
of residual errors when correcting for 
detector sensitivity to optical photons (see text).
Optical contamination effects
likewise cause slight
mismatches in energies of intrinsic detector features such
as the Au-M complex (2.2 keV), Ni-K (7.5 keV),
and Au-L$\alpha$ (9.7 keV).
There are $\sim$1300 counts in the O-K line.
}
\label{fig:brightspec}
\end{figure}

\begin{figure*}[t]
\vspace{5mm}
\epsscale{0.48}
\rotatebox{270}{\plotone{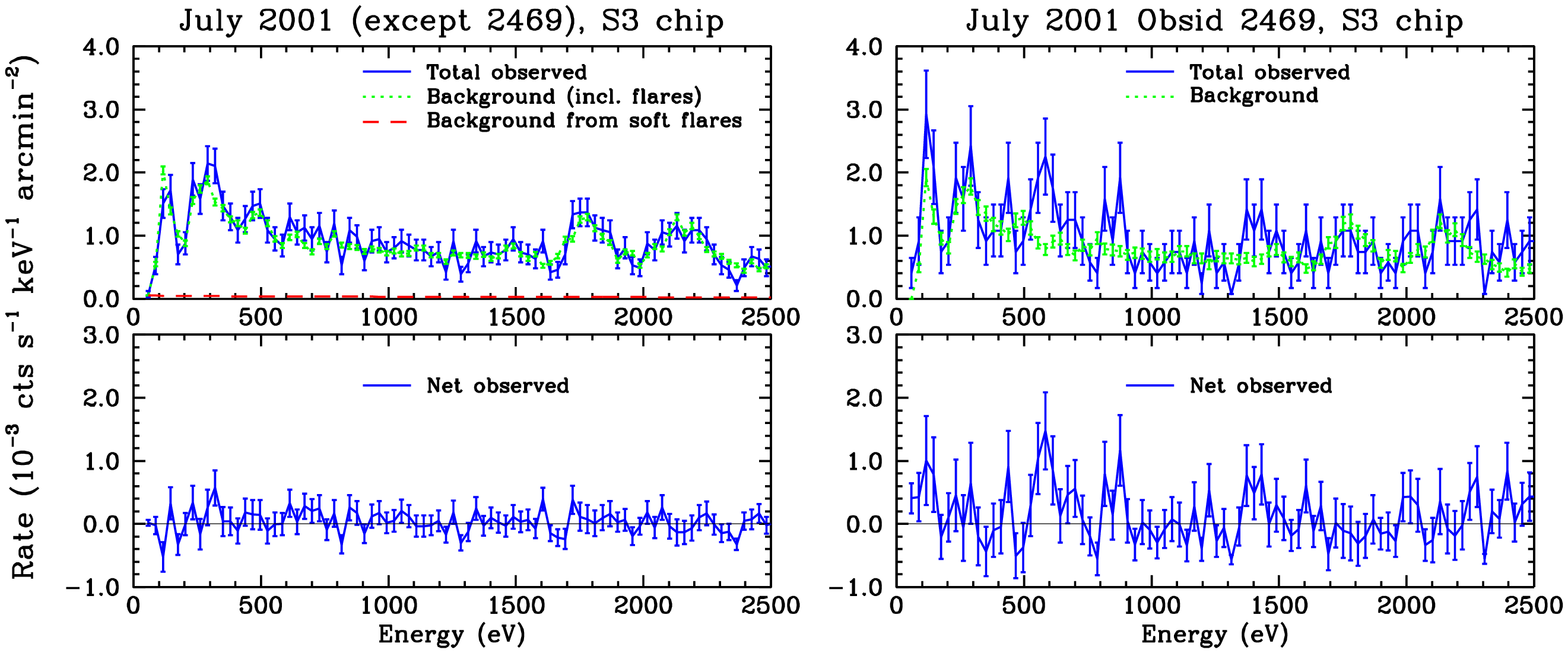}}

\vspace{5mm}
\rotatebox{270}{\plotone{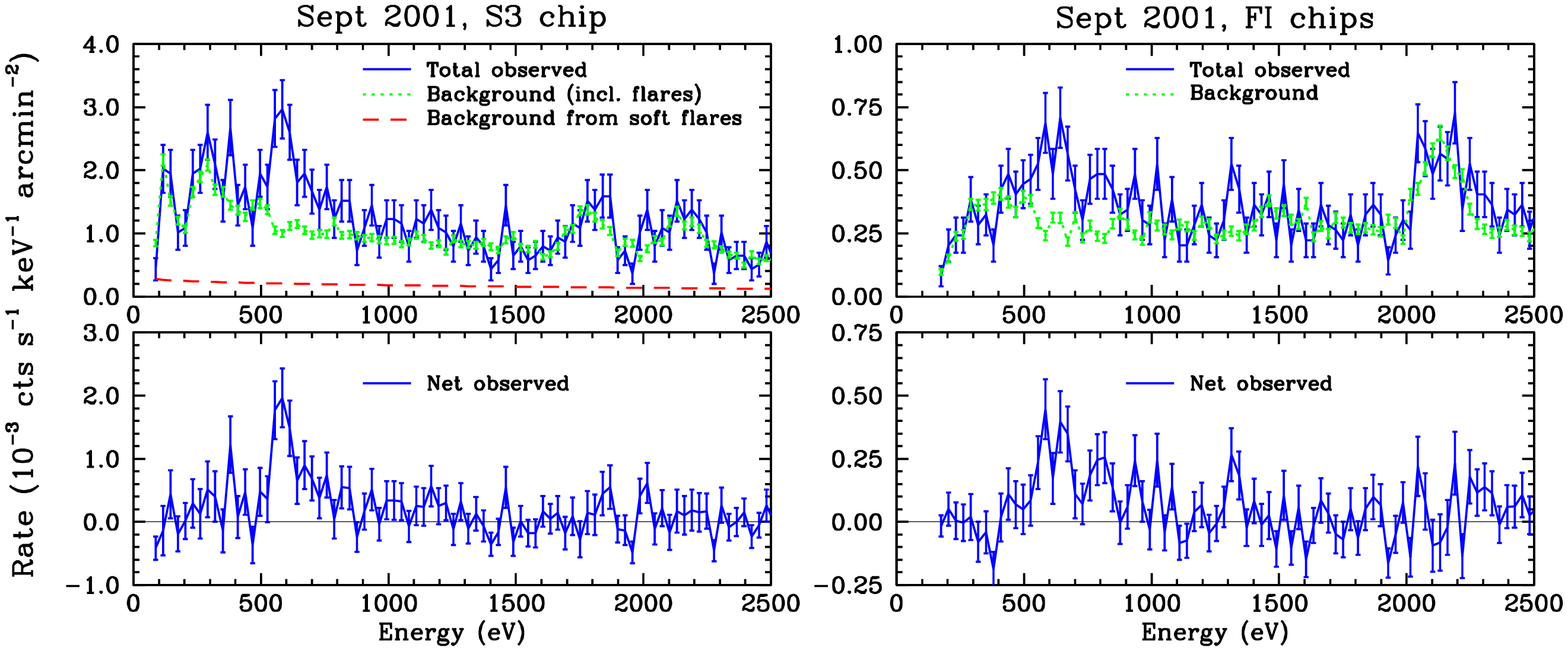}}
\vspace{4mm}

\caption{
Observed and background-subtracted spectra from the
dark side of the Moon, with 2-channel (29.2 eV) binning.
In the July S3 data (top four panels), 
no excess emission is obvious except in ObsID 2469.
Features around 1750 eV (Si-K) and 2150 eV (Au-M)
are from particle-induced fluorescence of the detector assembly.
Bottom four panels show S3 and combined FI spectra from
the three September ObsIDs with the strongest
emission excesses.
Oxygen emission from
charge transfer is clearly seen in both spectra, and energy resolution in
the FI chips is sufficient that \iolya\ is
largely resolved from \ioka.
High-$n$ \ion{O}{8} Lyman lines are also apparent in the FI spectrum,
along with what is likely \imgka\ around 1340 eV.
}
\label{fig:mainfig}
\end{figure*}

One last complication is the effect of optical contamination on the
energy calibration.  Although raw energy offsets were
removed from the data during the bias-correction process, 
more subtle effects remained, mostly
related to charge transfer inefficiency.  Optical-leak events 
partially fill the charge traps in the CCDs, thus reducing the net CTI.
When standard CTI corrections were applied to the data 
to improve the energy resolution
of the FI chips, this overcorrected and pushed energies too high
for data affected by the optical leak.
The effect varies between and within chips based on optical exposure, 
but we can place an upper
limit on it by examining the bright-Moon data, which are most affected.
Fig.~\ref{fig:brightspec} 
shows the spectrum of the combined July bright-Moon
data, which come from the I2 and I3 chips.  
The $K$-shell fluorescence lines of O, Mg, Al, and Si are
easily identified, and we find a positive offset of $50\pm5$ eV
from their true values of 525, 1254, 1487, and 1740 eV, respectively.
Bright-side data from ObsID 2469 I3 were excluded because they showed an offset
of roughly 80 eV, with a distorted shape for the O peak; 
to be conservative, we excluded the
corresponding dark-side data from further consideration as well.
As noted before, I2 data from ObsID 2469 were already excluded
because of their much larger optical contamination.

Energy offsets in the dark-side and CXB data should be much smaller,
particularly for the S chips, which 
did not image the bright side of the Moon.
Judging from the positions of weak fluorescence lines 
present in the detector background, and the agreement of
astrophysical line positions in the FI and S3 spectra
with each other and theoretical models (see \S\ref{sec:modelspec}), 
the dark-side
energy errors indeed appear to be negligible.


\section{RESULTS}
\label{sec:results}

\subsection{Dark-Side Spectra}
\label{sec:resultspectra}

X-ray spectra were created from the event files using CIAO {\tt dmextract}.
Data from the three FI chips (I2, I3, and S2), which have lower QE
than the BI chips below 1 keV, were always combined in
order to improve statistics.
Scaled background spectra,
with soft-flare corrections as needed, were created for each
observed dark-side spectrum and subtracted to reveal any
excess X-ray emission.  

Summing all the July data apart from ObsID 2469 reveals
no significant excesses in either the S3 or combined FI spectra
(see top half of Fig.~\ref{fig:mainfig}).
The S3 spectrum from ObsID 2469, however,
has a noticeable emission feature (more than $3\sigma$) at $\sim600$ eV.
The corresponding FI spectrum for ObsID 2469 has too few counts to be
used for corroboration as it includes only data from
the S2 chip (because of the optical leaks in I2 and I3).

Much stronger evidence for excess emission near 600 eV appears in the
September dark-Moon spectra, in all chips (bottom half of
Fig.~\ref{fig:mainfig}).  
It is immediately obvious that this can not be 
particle- or photon-induced O fluorescence,
which would occur in a single line at 525 eV, nor is it
electron-impact continuum
as posited by \citet{cit:schmitt1991}.

To assess the significance of any excesses,
we selected three energy ranges for statistical study
with the {\it a priori} assumption that solar-wind
charge transfer is the source of the emission (\S\ref{sec:interp}).
Each range (311--511 eV, 511--711 eV, and 716--886 eV)
was chosen to extend $\sim50$ eV below and above
the strongest CT lines expected within each range (\S\ref{sec:modelspec}).
For comparison, the 886--986 eV band was also studied
(in which we might hope to see \ineka\ at 905--922 eV), 
along with four 200-eV-wide bands from 1000 to 1800 eV.
The most important results are shown
in Table~\ref{table:ratestats}, with excesses of more than $2.5\sigma$
shown in bold.

\begin{deluxetable*}{cccccccccccccc}
\tablecaption{Net Emission Within Selected Energy Bands \label{table:ratestats}}
\tablehead{
        &       & & \multicolumn{2}{c}{311--511 eV} &   & \multicolumn{2}{c}{511--716 eV} &     & \multicolumn{2}{c}{716--886 eV} &     & \multicolumn{2}{c}{886--986 eV}       \\
\cline{4-5}     \cline{7-8}     \cline{10-11}   \cline{13-14}   \\
&       & & Counting & Signif.      & & Counting & Signif.   & & Counting & Signif.       & & Counting & Signif.        \\
Chips   & ObsIDs& & Rate\tablenotemark{a}& ($\sigma$) & & Rate\tablenotemark{a} & ($\sigma$) & & Rate\tablenotemark{a} & ($\sigma$) & & Rate\tablenotemark{a}   & ($\sigma$)     
}
\startdata
FIs\tablenotemark{b}  
    & 2469& & $  2\pm 19$   & 0.1 & & $         25 \pm 20 $ &      1.2  & & $          6\pm 16 $ &      0.4     & & $ 25\pm 16$ & 1.5   \\
FIs & all July\tablenotemark{c} &
            & $ -1\pm  6$   &-0.2 & & $          3 \pm  6 $ &      0.5  & & $          7\pm  5 $ &      1.5     & & $  3\pm  4$ & 0.7   \\
\\
FIs & 2468& & $  2\pm 12$   & 0.1 & & $\mathbf{ 65 \pm 14}$ & {\bf 4.5} & & $\mathbf{ 32\pm 12}$ & {\bf 2.8}    & & $ -1\pm  7$ &-0.1   \\
FIs & 3368& & $-11\pm 13$   &-0.9 & & $\mathbf{ 76 \pm 18}$ & {\bf 4.2} & & $         22\pm 13 $ &      1.7     & & $ 18\pm 11$ & 1.7   \\
FIs & 3370& & $ -5\pm  9$   &-0.5 & & $         12 \pm  9 $ &      1.3  & & $         18\pm  9 $ &      2.1     & & $  2\pm  6$ & 0.3   \\
FIs & 3371& & $  1\pm 11$   & 0.1 & & $         26 \pm 12 $ &      2.2  & & $         22\pm 10 $ &      2.1     & & $ 15\pm  8$ & 1.9   \\
FIs & 3 bright\tablenotemark{d}&        
            & $ -2\pm  7$   &-0.3 & & $\mathbf{ 52 \pm  8}$ & {\bf 6.3} & & $\mathbf{ 26\pm  7}$ & {\bf 3.9}    & & $ 10\pm  5$ & 2.1   \\
\\
S3  & 2469& & $-36\pm 34$   &-1.1 & & $\mathbf{126 \pm 39}$ & {\bf 3.2} & & $         29\pm 30 $ &      1.0     & & $-19\pm 17$ &-1.1   \\
S3  &quiet July\tablenotemark{e}& 
            & $ 30\pm 18$   & 1.7 & & $         22 \pm 15 $ &      1.4  & & $         13\pm 13 $ &      1.0     & & $  6\pm 10$ & 0.7   \\
S3  & all July\tablenotemark{c}&        
            & $ 22\pm 16$   & 1.4 & & $\mathbf{ 43 \pm 14}$ & {\bf 3.0} & & $         18\pm 12 $ &      1.5     & & $  4\pm 8$  & 0.4   \\
\\
S3  & 2468& & $ 51\pm 31$   & 1.7 & & $\mathbf{218 \pm 33}$ & {\bf 6.6} & & $         27\pm 23 $ &      1.2     & & $  9\pm 16$ & 0.5   \\
S3  & 3368& & $ 54\pm 54$   & 1.0 & & $\mathbf{256 \pm 61}$ & {\bf 4.2} & & $         18\pm 40 $ &      0.4     & & $ 11\pm 30$ & 0.4   \\
S3  & 3370& & $ 58\pm 37$   & 1.5 & & $         77 \pm 34 $ &      2.3  & & $        -21\pm 24 $ &     -0.9     & & $-15\pm 17$ &-0.9   \\
S3  & 3371& & $ 26\pm 49$   & 0.5 & & $\mathbf{247 \pm 60}$ & {\bf 4.1} & & $\mathbf{214\pm 54}$ & {\bf 4.0}    & & $ 72\pm 35$ & 2.0   \\
S3  & 3 bright\tablenotemark{d}&
            & $ 63\pm 28$   & 2.2 & & $\mathbf{228 \pm 31}$ & {\bf 7.3} & & $\mathbf{ 58\pm 23}$ & {\bf 2.6}    & & $ 18\pm 16$ & ~~~~~~1.2
\enddata
%
%
\tablecomments{Rate excesses of more than 2.5$\sigma$ are shown in bold.
``FIs'' means the I2, I3, and S2 chips in combination.\\
$^a$~Units of $10^{-6}$ cts s$^{-1}$ arcmin$^{-2}$.~
$^b$~Chip S2 only; data from I2 and I3 are unusable.~
$^c$~ObsIDs 2469, 2487, 2488, 2489, 2490, and 2493.~
$^d$~The ``bright-September'' ObsIDs: 2468, 3368, and 3371.~
$^e$~All July ObsIDs except 2469.
}
\end{deluxetable*}

Emission in the
511--711 eV range has an excess of more than $4\sigma$
in three of the four S3 spectra from September,
and $2.3\sigma$ in the other (ObsID 3370).
The same pattern (most significant excess in ObsID 2468,
least in ObsID 3370) holds for the combined FI spectra.
When spectra from the three ObsIDs showing the largest excesses 
(2468, 3368, and 3371, henceforth referred to as the 
``bright-September'' ObsIDs) were summed,
the feature significance was more than $6\sigma$
in both the S3 and FI spectra (see Table~\ref{table:ratestats}).
The ratio of net counting rates for S3 and the FI chips
($4.3 \pm 0.9$) also matches well with the ratio
of those chips' effective areas in that energy range,
consistent with this being an X-ray signal from the sky.
Results from the S1 chip were consistent with those from S3,
but with lower significance because of the S1's much
shorter dark-Moon exposure times and somewhat higher background;
we do not discuss S1 dark-side results further.

The 716--886 eV range, which we expect to contain
\iolyb\ and \iolyd\ emission,
also showed significant excesses in the
S3 and combined FI spectra for the bright-September ObsIDs
($\sim2.5\sigma$ and $\sim 4 \sigma$, respectively)
with excesses in individual ObsIDs roughly following the
time pattern seen in the 511--711-eV band.
The S3 spectrum for ObsID 3371 stands out with a $4\sigma$ excess.
The same ObsID also has
$2\sigma$ excesses in the 886--986 eV range in both the
S3 and FI spectra.  
This energy range, like the 716--886 eV range,
contains CT emission lines (\ineka\ at 905--922 eV) 
from an ion which is only
abundant when the solar wind is especially highly ionized.
As we will discuss in \S\ref{sec:windadjust}, there is evidence
for such a situation during ObsID 3371.

Above 1000 eV, no significant excesses
were seen for any of the ObsID combinations listed
in Table~\ref{table:ratestats} with 
the possible exception of 1200--1400 eV, which
had a $2.9\sigma$ excess in the bright-September FI spectrum
(and a $1.0\sigma$ excess in the S3 spectrum).
Again, 3371 was the individual ObsID recording the
largest excesses in the FI ($1.8\sigma$) and S3 ($1.3\sigma$) spectra.
Although the evidence is not compelling,
we believe that the observed excess
probably represents a detection of 
He-like \imgka\ ($\sim 1340$ eV).

In the 311--511 eV range where \ion{C}{6} Lyman emission might be detectable,
only a $2.2\sigma$ excess appears in the S3 data.
Over the full range of O emission (511--886 eV), 
which is relevant to the discussion in \S\ref{sec:interp},
the bright-September S3 and FI spectra both have an excess of $7.4\sigma$,
with net rates of $287 \pm 39$ and $78 \pm 11 \times 10^{-6}$
cts s$^{-1}$ arcmin$^{-2}$, respectively.
The summed July S3 data have a rate of
$62 \pm 19 \times 10^{-6}$ cts s$^{-1}$ arcmin$^{-2}$ between 511 and 886 eV.
If ObsID 2469 is excluded because of its obviously stronger O emission,
the rest of the July S3 data have a statistically insignificant
$1.7\sigma$ excess with a rate of $34 \pm 20 \times 10^{-6}$ 
cts s$^{-1}$ arcmin$^{-2}$.

(As an aside, we note that the S3 Moon spectrum from July,
combining data from all six ObsIDs, was
used as a measure of the detector background by \citet{cit:maxim2002}.
Even with the ObsID-2469 excess, the oxygen emission in that background 
is much less than the sky-background emission discussed in that paper,
and so the authors' results are not significantly affected.)

\subsection{Comparison with CXB}
\label{sec:resultscxb}

If the dark-side emission arises between the Earth and Moon,
then it must also be present at the same level 
on the bright side and in 
the cosmic X-ray background beyond the Moon's limb.
Unfortunately, the bright side was observed using the I2 and I3 chips
and only in July, 
when the dark-side emission was barely detectable even with the
more sensitive S3 chip.  Such a weak signal would in any
case be swamped by the bright-side fluorescence X-rays.

\begin{figure}[b]
\includegraphics[scale=0.44,bb=24 284 590 750,clip]{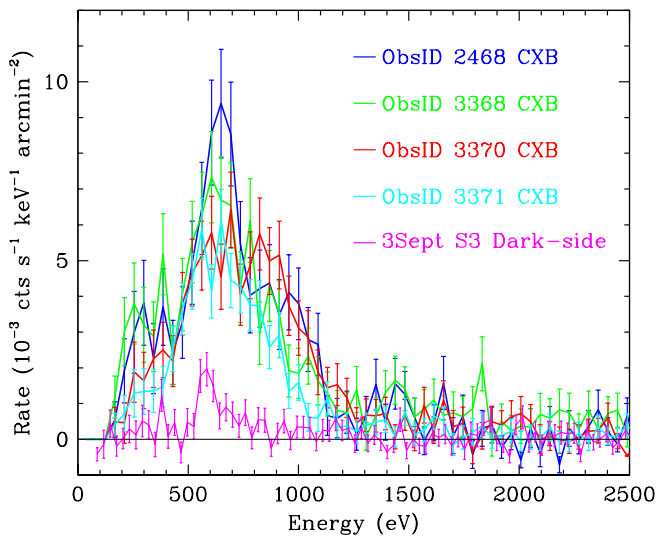}

\vspace{1mm}
\caption{
Background-subtracted CXB spectra from the S1 chip
for all four September observations,
with 3-channel binning (43.8 eV) for clarity.
Spectra are not directly comparable
because they were collected from slightly different regions of the sky
(see Table~\ref{table:cxb}),
all of which are quite bright.
The bright-September dark-side S3 spectrum is shown for comparison.
}
\label{fig:apjCXB}
\end{figure}

Spectra of the CXB
were obtained only in September, primarily
by the S1 chip (see Table~\ref{table:exposures}) 
which is back-illuminated like S3 and
has similar quantum efficiency.  
In Fig.~\ref{fig:apjCXB}, which shows CXB S1 spectra from all
four September observations, it is apparent that
the CXB is much brighter than the dark-Moon signal.
In fact, this is one of the brightest regions of the sky,
with a complex spatial structure;
see Table~\ref{table:cxb}, which lists
the centroid of each CXB extraction region 
and the corresponding 
R45 (3/4-keV-band) RASS rate.\footnote{%
Background maps are available online at
http://www.xray.mpe.mpg.de/rosat/survey/sxrb/12/ass.html.
An X-ray Background Tool is also available at
http://heasarc.gsfc.nasa.gov/cgi-bin/Tools/xraybg/xraybg.pl.
Given the limited resolution of the PSPC,
data are usually divided into three energy bands:
R12 (a.k.a. the 1/4-keV band, effectively defined on the high-energy end
by the C absorption edge at 0.284 keV),
R45 (3/4-keV band; roughly 0.4--1.0 keV),
and R67 (1.5-keV band; roughly 1.0-2.0 keV).
}
The typical SXRB recorded in the RASS is roughly one-quarter
as bright, comparable to the bright-September dark-Moon brightness.

\tabletypesize{\small}
\begin{deluxetable}{cccc}[!b]
\tablecaption{\rosat\ All-Sky Survey Background Rates
 for CXB Regions\label{table:cxb}}
\tablehead{
                &                               &
                & \colhead{R45 Rate} \\
\colhead{ObsID} & \colhead{Centroid RA}         & \colhead{Centroid Dec}  
                & \colhead{($10^{-6}$ cts s$^{-1}$ arcmin$^{-2}$)}
}
\startdata
2468    & $15^{h}41^{m}44^{s}$  & $-37^{\circ}00^{\prime}30^{\prime\prime}$     & $543.7 \pm 103.2$ \\
3368    & $15^{h}41^{m}29^{s}$  & $-37^{\circ}15^{\prime}10^{\prime\prime}$     & $483.3 \pm 100.3$ \\
3370    & $15^{h}41^{m}18^{s}$  & $-37^{\circ}29^{\prime}10^{\prime\prime}$     & $549.0 \pm 107.1$ \\
3371    & $15^{h}41^{m}30^{s}$  & $-37^{\circ}41^{\prime}20^{\prime\prime}$     & $468.7 \pm 100.5$
\enddata
\tablecomments{R45 band is defined as PI channels 52--90, corresponding
to approximately 0.4--1.0 keV.
\rosat\ rates were found using the HEASARC X-Ray Background Tool version 2.1
(http://heasarc.gsfc.nasa.gov/Tools)
with a cone radius of 0.1 degrees.
}
\end{deluxetable}

One can also see that
emission around 600 eV is strongest for ObsID 2468,
which is also the ObsID that shows the most significant
dark-side X-ray emission.
While this correlation is suggestive,
direct comparisons among the CXB spectra are
not possible because they were taken from slightly different
regions of the sky (because of the Moon's motion),
nor can the CXB data be adequately normalized using
\rosat\ All-Sky Survey background rates because
statistical uncertainties are too large.

\section{INTERPRETATION: GEOCORONAL CHARGE TRANSFER}
\label{sec:interp}

The primary result of our analysis is that highly-significant time-variable
emission is seen looking toward the dark side of the Moon
at energies between 500 and 900 eV.
As we discuss in this section, 
the observed spectrum, intensity, and temporal behavior
can all be explained by charge transfer between the solar wind
and the Earth's outer atmosphere.

Charge transfer is the radiationless collisional
transfer of one or more electrons from a neutral atom or molecule
to an ion.  When the recipient ion is highly charged, it is left
in a high-$n$ excited state which then decays via single or 
sequential radiative transitions. 
(The ion may also autoionize if multiple electrons are transferred
from the neutral.)
In geocoronal CT, the neutral gas is atomic H in the Earth's
outer atmosphere extending tens of thousands of km into space,
the X-ray emitting ions are heavy elements such as C, O, and Ne
in the solar wind, and the collisions take place
outside the magnetosphere, into which the wind
particles can not penetrate.  At X-ray energies, most of the emission
comes from hydrogenic and He-like C and O ions because of
their relatively high abundance.

\subsection{Model Spectra}
\label{sec:modelspec}

The equation for the CT emissivity of line $l$ from ion $i$
can be written as
\begin{equation}
\epsilon_{il} = 
        v_{c}
        n_{n} 
        n_{i}
        y_{il} 
        \sigma_{i}\;\;
\mathrm{phot} \; \mathrm{s}^{-1}\mathrm{cm}^{-3},
\end{equation}
where $v_{c}$ is the collision velocity
(effectively the solar wind velocity),
$n_{n}$ is the neutral species density,
$n_{i}$ is the relevant ion density,
$y_{il}$ is the net line emission yield per CT-excited ion,
and $\sigma_{i}$ is the
total CT cross section for ion $i$.

\tabletypesize{\small}
\begin{deluxetable}{lccc}
\tablecaption{Slow Solar Wind Ion Abundances 
and Cross Sections \label{table:ions}}
\tablewidth{0.9\linewidth}
\tablehead{
                & \colhead{Abund.} & \colhead{$\sigma_{CT}$ with H} & \colhead{$\sigma_{CT}$ with He} \\
\colhead{Ion}   & \colhead{rel. to O}   & \colhead{($10^{-15}$ cm$^{2}$)}& \colhead{($10^{-15}$ cm$^{2}$)}
}
\startdata
~~\ion{C}{6}      & 0.21  & 1.03  & 1.3   \\
~~\ion{C}{7}      & 0.32  & 4.16  & 1.5   \\
~~\ion{N}{7}      & 0.06  & 3.71  & 1.7   \\
~~\ion{N}{8}      & 0.006 & 5.67  & 2.0   \\
~~\ion{O}{7}      & 0.73  & 3.67  & 1.1   \\
~~\ion{O}{8}      & 0.20  & 3.4   & 1.8   \\
~~\ion{O}{9}      & 0.07  & 5.65  & 2.8   \\
~~\ion{Ne}{9}     & 0.084 & 3.7   & 1.5   \\
~~\ion{Ne}{10}    & 0.004 & 5.2   & 2.4
\enddata
\tablecomments{Relative element abundances: C/O = 0.67, N/O = 0.0785,
Ne/O = 0.088.
}
\end{deluxetable}

As described in the review by \citet{cit:smith2003},
solar wind velocity and ionization level are closely correlated,
and can be used to divide the solar wind into two main types:
a ``slow'' highly ionized wind ($v_{c} \sim 400$ km s$^{-1}$),
and a ``fast'' less ionized wind ($v_{c} \sim 700$ km s$^{-1}$).
During solar minimum, the slow wind is found near the equatorial plane
(solar latitudes between roughly $-20^{\circ}$ and $+20^{\circ}$),
while the fast wind dominates at higher latitudes.
During solar maximum, the slow wind extends to higher latitudes,
but there is significant mixing of the two components.
Average slow-wind ion abundances relative to oxygen, 
taken from \citet{cit:schwadron2000}, 
are listed in Table~\ref{table:ions}, along with cross sections
for CT with neutral H and He.

\tabletypesize{\small}
\begin{deluxetable}{lcc}
\tablecaption{Major Solar CT Emission Lines \label{table:lines}}
\tablewidth{0.9\linewidth}
\tablehead{
                & \colhead{Energy} &  \\
\colhead{Line}  & \colhead{(eV)}   & \colhead{Line Yield}
}
\startdata
~~\icka\tablenotemark{a}    & 299, 304, 308       & 0.899    \\
~~\iclya                          & 368           & 0.650    \\
~~\inka\tablenotemark{a}    & 420, 426, 431       & 0.872    \\
~~\iclyb                          & 436           & 0.108    \\
~~\iclyg                          & 459           & 0.165    \\
~~\ioka\tablenotemark{a}    & 561, 569, 574       & 0.865    \\
~~\iolya                          & 654           & 0.707    \\
~~\iokb                           & 666           & 0.121    \\
~~\iolyb                          & 775           & 0.091    \\
~~\iolyg                          & 817           & 0.033    \\
~~\iolyd                          & 836           & 0.103    \\
~~\iolye                          & 847           & 0.030    \\
~~\ineka\tablenotemark{a}   & 905, 916, 922       & 0.887
\enddata

\tablenotetext{a}{The He-like K$\alpha$ complex consists of three lines:
forbidden ($1s2s$ $^{3}S_{1} \rightarrow 1s^{2}$ $^{1}S_{0}$), 
intercombination ($1s2p$ $^{3}P_{1,2} \rightarrow 1s^{2}$ $^{1}S_{0}$), and
resonance ($1s2p$ $^{1}P_{1} \rightarrow 1s^{2}$ $^{1}S_{0}$).
For charge transfer, the forbidden line (lowest energy) dominates.
}
\end{deluxetable}

A great deal of physics is contained within 
$y_{il} \sigma_{i}$, as it includes
the initial quantum-sublevel population distribution of the ion
immediately following electron capture,
and then the branching ratios from all those levels during
the subsequent radiative cascades.
Cross sections for CT of highly charged C, N, O, and Ne ions 
with atomic H and associated radiative branching ratios 
are taken from 
\citet{cit:harel1998}, 
\citet{cit:greenwood2001}, 
\citet{cit:rigazio2002},
\citet{cit:johnson2000},
and related references in \citet{cit:khar2000,cit:khar2001}.
Total CT cross sections for all ions are a few $\times 10^{-15}$ cm$^{-2}$,
with uncertainties of typically 30\%, and
are fairly constant as a function of collision velocity near
400 km s$^{-1}$.
Cross sections for electron capture into
individual sublevels have larger errors,
which are the major contributors to uncertainties in the line yields,
particularly for H-like ions.
As can be seen from the line yields listed in Table~\ref{table:lines},
emission from He-like ions is predominantly ($\sim$90\%)
in the form of K$\alpha$ ($n = 2 \rightarrow 1$) photons,
while H-like emission is split more evenly between
Ly$\alpha$ and the higher-$n$ transitions 
(e.g., Ly$\gamma$ and Ly$\delta$).
The unusual strength of the high-$n$ Lyman lines
is a unique signature of CT 
which can not be reproduced by thermal plasmas.

\begin{figure}[b]
\includegraphics[angle=270,scale=0.53,bb=60 85 364 548]{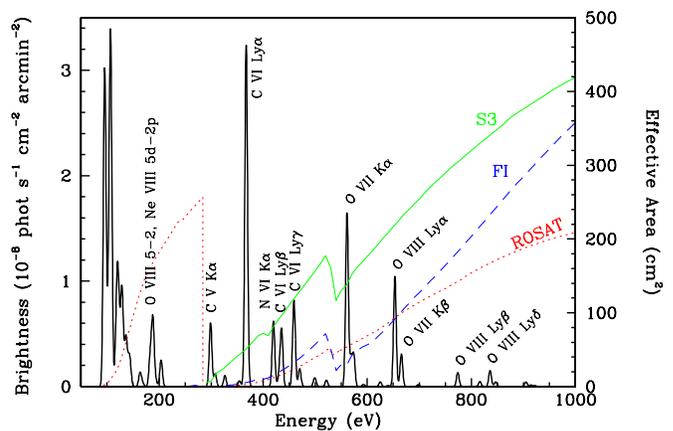}

\vspace{1mm}
\caption{
Model geocoronal X-ray spectrum
and effective areas for \chandra\ and \rosat.
Brightness is for an observation through the magnetosphere
flank assuming average slow-solar-wind parameters, 
plotted with FWHM energy resolution of 6 eV.
The cluster of lines below 150 eV comprises mostly
$n=4,5 \rightarrow 2$ transitions in \ion{O}{6} and \ion{O}{7}.
}
\label{fig:modelspec}
\end{figure}

For comparison with work by \citet{cit:cravens2001}
and \citet{cit:robertson2003},
we calculate the value of
$\alpha = \sum_{il} (n_{i}/n_{p}) y_{il} \sigma_{i} E_{il}$,
where $n_{p}$ is the solar-wind proton density,
$E_{il}$ is the line energy, and the sum is over all CT lines
from 95 eV (the lower limit of \rosat's energy range) to 1000 eV.
\citet{cit:robertson2003} estimate that
$\alpha$ equals $6 \times 10^{-16}$ eV cm$^{2}$ for CT
with H and $3 \times 10^{-16}$ eV cm$^{2}$ for He.
We derive $9.82 \times 10^{-16}$ and $4.56 \times 10^{-16}$
eV cm$^{2}$, respectively, or 
$8.13 \times 10^{-16}$ and $3.92 \times 10^{-16}$ eV cm$^{2}$
for energies above 180 eV, where the \rosat\ PSPC 
effective area becomes appreciable (see Fig.~\ref{fig:modelspec}).
Whatever its value, a global parameter such
as $\alpha$ is insufficient for the work described here;
when analyzing data with at
least moderate energy resolution it is necessary to
create a line emission model.


\begin{figure*}[t]
\centerline{\epsscale{1.0}\rotatebox{0}{\plotone{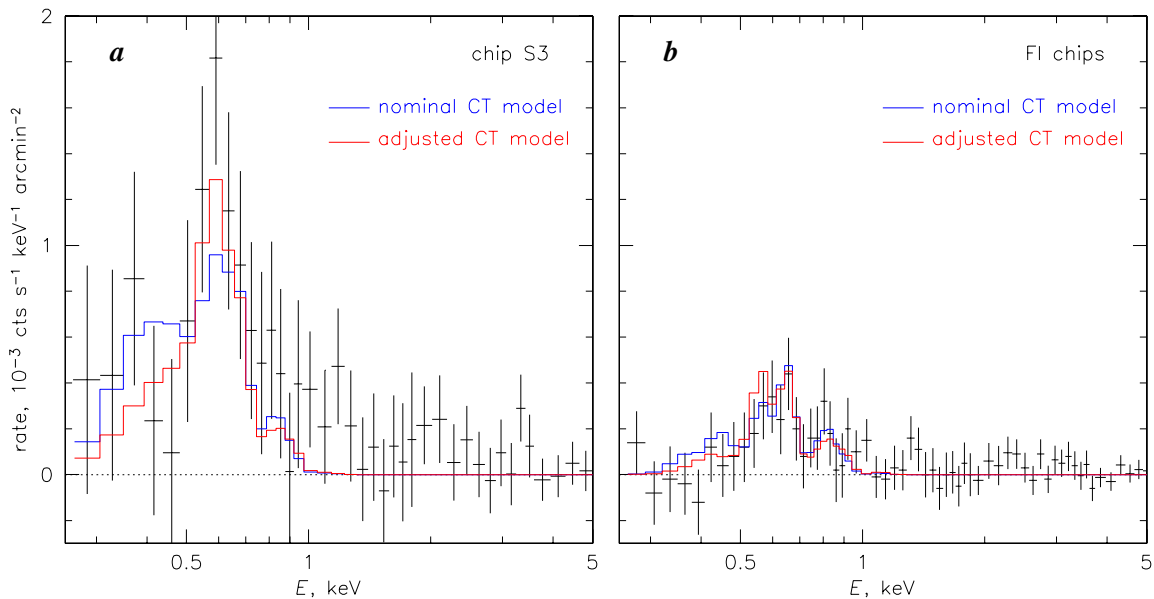}}}
\caption{
Background-subtracted September spectra (3 brightest ObsIDs)
and charge transfer model fits using four composite lines.
The same model, including normalization, is used to fit
the S3 and FI spectra.
The nominal fit uses the average (baseline) solar-wind parameters
discussed in the text.
In the adjusted fit, the relative
\ion{C}{6} vs \ioka\ emission is 1/6 of the nominal case,
and \ion{O}{8} emission is reduced by half.
Results for the adjusted fit are listed in Table~\ref{table:fitbright}.  
}
\label{fig:fitbright}
\end{figure*}

\tabletypesize{\small}
\begin{deluxetable}{cclccc}
\tablecaption{Fit Results versus 
        Model Line Brightness\label{table:fitbright}}
\tablehead{
\colhead{Energy}        
        & \colhead{Range\tablenotemark{a}}      
                        & \colhead{Major}       
                                        &               &       &       \\
\colhead{(eV)}  
        & \colhead{(eV)}        
                        & \colhead{Lines}       
                                        & \colhead{$B_{fit}$\tablenotemark{b}}
                                        & \colhead{$B_{model}$\tablenotemark{c}}
                                        &\colhead{$\frac{B_{fit}}{B_{model}}$}
}
\startdata
440     & 367--470      &C Ly$\alpha$-$\delta$  & 32 ( 0--76)   & 5.58  &  6  \\
570     & 561--574      &O K$\alpha$            & 85 (55--115)  & 2.38  & 36  \\
660     & 654--666      &O Ly$\alpha$, K$\beta$ & 38 (23--52)   & 0.96  & 40  \\
810     & 775--847      &O Ly$\beta$-$\epsilon$ & 15 ( 9--22)   & 0.23  & 66  \\
1340    & $\cdots$      &Mg K$\alpha$   &  4 ( 1-- 7)   & $\cdots$ & $\cdots$
\enddata
\tablenotetext{a}{Range of energies for grouped CT-model lines.}
\tablenotetext{b}{Fit results for the 3 bright-September ObsIDs,
        in units of $10^{-8}$ phot s$^{-1}$ cm$^{-2}$ arcmin$^{-2}$.
        Parentheses denote 90\% confidence limits.}
\tablenotetext{c}{Model line brightnesses are calculated using the average
        slow-solar-wind parameters listed in the text, in units
        of $10^{-8}$ phot s$^{-1}$ cm$^{-2}$ arcmin$^{-2}$.
        Detailed model predictions for the \imgka\ line at 1340 eV are
        not available, but its normal brightness is negligibly small.}
\end{deluxetable}

Our resulting model spectrum for geocoronal CT is shown in 
Fig.~\ref{fig:modelspec}.
That model was then used to simultaneously fit the 
bright-September S3 and FI data
between 250 and 5000 eV
using their associated ACIS response functions (see \S\ref{sec:spatial}).
Given the limited statistics ($\sim700$ counts attributable to
X-ray emission in the S3 data, and fewer in the FI spectra),
we grouped the model emission into four ``lines'' at 
440, 570, 660, and 810 eV;
the 440-eV and 810-eV lines are modeled as finite-width
Gaussians since they represent several lines spread
over a 100-eV and 60-eV range, respectively.

Because of the poor statistics, the only free parameter is
the normalization.
As can be seen in Fig.~\ref{fig:fitbright},
overall agreement between the shape of the model spectrum and observation
is good, particularly for the O lines
which are the most prominent features.
The fit shows that carbon emission is overpredicted, however,
which may be because
of atypical ion abundances, errors in CT cross sections
and line yields, or
uncertainties in the QE of the ACIS detectors
near the extreme limits of their effective energy ranges.
Better agreement is obtained when 
relative ion abundances are adjusted;
in the adjusted fit, we reduce the abundances (relative to He-like \ion{O}{7})
of H-like \ion{C}{6} and \ion{O}{8} by factors of 6 and 2, respectively.
When a line is added to fit the putative \imgka\ feature at 1340 eV,
the $F$-test significance is 99.6\%.

In fitting the FI and S3 spectra simultaneously, we are
implicitly assuming that they recorded the same emission.
This is not strictly true because of differing exposures for
each chip/ObsID combination, and in fact, if the
S3 and FI spectra are fit separately, the S3 brightness comes
out somewhat higher than for the FI data.
Although within statistical uncertainties, this difference is consistent
with the sense of suspected errors in the quantum efficiency
of the ACIS chips.

Line brightnesses from the adjusted-abundances fit are 
listed in Table~\ref{table:fitbright};
note that \ioka\ and \iolya\ are essentially unresolved
in the S3 data so their combined brightness is more reliable
than the individual values.
We also list the corresponding model predictions for an average solar wind
(which will be derived in \S\ref{sec:predrates}),
which are more than an order of magnitude smaller than observed.
As we will discuss in \S\ref{sec:windadjust}, based on available
solar-wind data, there is
good reason to expect a much higher-than-average
geocoronal emission level
during the September observations,
as well as relatively weak \ion{C}{6} emission.

We lastly note that the bright-September S3 spectrum is
remarkably similar in shape (and normalization)
to the diffuse X-ray background spectra
reported by \citet{cit:maxim2002} for ObsIDs 3013 and 3419,
which observed regions of the sky removed from any
bright Galactic structures (see their Fig.~14).  
Those spectra were fit with a MEKAL thermal model, in
both cases yielding a temperature of $T\sim0.2$ keV.
We obtain the same result when fitting that model to 
our bright-September data.

\subsection{Predicted Intensities for Average Solar Wind}
\label{sec:predrates}

\begin{figure}[t]
\vspace{2mm}
\epsscale{1.1}
\rotatebox{0}{\plotone{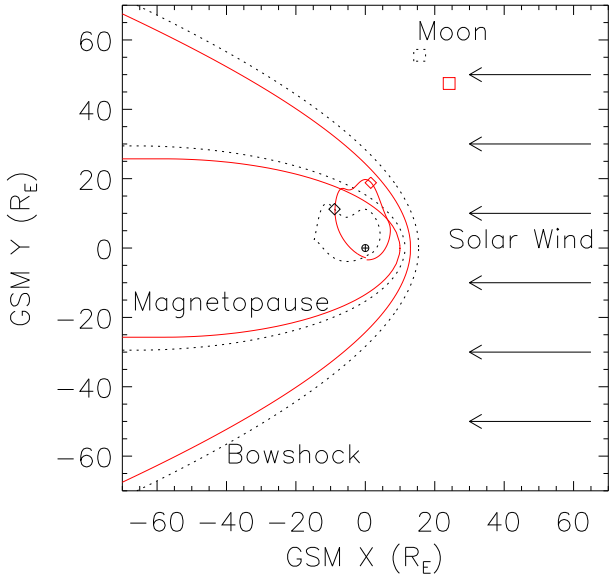}}

\vspace{1mm}
\caption{
Schematic of observation viewing geometry, with
dotted curves for the July observations and solid for September.
In Geocentric Solar Magnetospheric (GSM) coordinates,
Earth is at the origin, $X$ points toward the Sun,
and the $XZ$ plane contains the Earth's magnetic axis.
The $Y$ and $Z$ axes thus oscillate with a period of 24-hours,
which makes \chandra's orbit appear non-elliptical.
Positions of the magnetosphere and bowshock are
approximate and based on the models of
\citep{cit:tsyg1995,cit:tsyg1989} and 
\citep{cit:bennett1997}, respectively.
\chandra\ was barely inside the magnetopause in July,
despite its apparent position in this projection.
}
\label{fig:xy_viewing_geometry_v2}
\end{figure}

To compare predicted absolute line intensities with observed values,
we return to Eq.~1, and write the emissivity as
$\epsilon_{il} = 
        v_{c}
        n_{n} 
        n_{p} f_{O} f_{i}
        y_{il} 
        \sigma_{i}$,
where the ion density, $n_{i}$, is expressed 
as the solar-wind (proton) density, $n_{p}$, times the 
relative abundance of oxygen, $f_{O}$, times
the ion abundance relative to O, $f_{i}$.
As a baseline for comparison with specific \chandra\ observations,
we use average solar-wind parameters in the calculations
that follow.
As noted before, the average slow-wind velocity $v_{c}$ is 400 km s$^{-1}$.
Solar wind density is 7 cm$^{-3}$ at 1 AU,
the fractional O abundance is about 5.6$\times10^{-4}$
\citep{cit:schwadron2000},
and relative ion abundances are listed in Table~\ref{table:ions}.
We assume that the ion density inside the magnetosphere is zero,
and constant everywhere outside.
Using the undisturbed wind density is an approximation, 
as the wind sweeps around the Earth's magnetosphere in 
a wake-shaped structure (see Fig.~\ref{fig:xy_viewing_geometry_v2}), 
piling up on the leading edge of the bowshock,
with higher-density lower-velocity shocked ions
in the magnetosheath, but the resulting errors are comparable
to other uncertainties in our model.

The remaining factor in CT emissivity, and the one with
the largest uncertainty, is the neutral gas density.
We use the analytical approximation of \citet{cit:cravens2001},
based on Hodges' (1994) model of exospheric hydrogen, which is
that atomic H density falls off roughly as $(10R_{E}/r)^{3}$,
where $R_{E}$ is one Earth radius (6378 km) and $r$ is 
the geocentric distance.
On the leading edge of the magnetosphere, at $r\sim10R_{E}$,
$n_{n}$ is approximately 25 cm$^{-3}$.
On the flanks, the magnetosphere extends to $r\sim15R_{E}$.


The brightness of a line $l$
observed by \chandra\ 
while looking toward the Moon
is given by 
\begin{equation}
B_{il} = \frac{1}{4\pi} \int_{0}^{D_{Moon}} \epsilon_{il}(x) dx\;\;
        \mathrm{phot}\; \mathrm{s}^{-1}\mathrm{cm}^{-2} \mathrm{sr}^{-1}
\end{equation}
where $x$ is the distance from \chandra.
The neutral gas density, and therefore emissivity, is essentially
zero at the distance of the Moon so we can replace $D_{Moon}$ with
$\infty$ in the integral.
If we also approximate the look direction as being radially outward 
from Earth,
then $dx = dr$ and
\begin{eqnarray}
B_{il} & = &    \frac{1}{4\pi} \int_{r_{min}}^{\infty} n_{n} 
                n_{p} f_{O} f_{i} v_{c} y_{il} \sigma_{i} dr \\
      & = &     \frac{v_{c} n_{p} f_{O} f_{i} y_{il} 
                \sigma_{i} n_{n0}}{4\pi}    \int_{r_{min}}^{\infty} 
                \left( \frac{10R_{E}}{r} \right)^{3}dr          \\
      & = &     \frac{v_{c} n_{p} f_{O} f_{i} y_{il} 
                \sigma_{i} n_{n0}}{4\pi}
                5R_{E} \left( \frac{10R_{E}}{r_{min}} \right)^{2},
\end{eqnarray}
where $r_{min}$ is the geocentric distance to the edge of the magnetosphere
or the spacecraft's position, whichever is farther.
The brightness of line $l$ for average slow-wind parameters is then
\begin{equation}
B_{il} = 9.95 \times 10^{14}
                \left( \frac{10R_{E}}{r_{min}} \right)^{2}
                y_{il} f_{i} \sigma_{i}\;\;             
    \mathrm{phot}\; \mathrm{s}^{-1}\mathrm{cm}^{-2} \mathrm{sr}^{-1}.
\end{equation}

\chandra\ orbits between about $3R_{E}$ and $20R_{E}$,
and the Moon observations were all made near apogee, so 
$r_{min} = 20 R_{E}$.
(\chandra\ was in fact very slightly inside the magnetopause in
July, but the difference in $r_{min}$ is negligible.)
During both July and September, \chandra\ was looking toward
the Moon through the forward flanks pointing slightly toward the leading edge
(see Fig.~\ref{fig:xy_viewing_geometry_v2}),
and its orbit was inclined by $\sim40^{\circ}$.
A numerical integration taking into account the non-radial character
of \chandra's viewing angle results in a factor of $\sim 1.5$ increase 
over the radial approximation.
A further slight adjustment for the effects of the shock region
(guided by the results of \citet{cit:robertson2003grl})
brings the total increase to a factor of two,
so that the final prediction for the \chandra\ observations,
assuming a typical slow solar wind, is a line brightness of
$B_{il}=5.0 \times 10^{14} y_{il} f_{i} \sigma_{i}\;\;
    \mathrm{phot}\; \mathrm{s}^{-1}\mathrm{cm}^{-2} \mathrm{sr}^{-1}$.
The $B_{model}$ values listed in Table~\ref{table:fitbright}
are the sums of $B_{il}$ for lines within each line-group.

It is often more convenient to compare predicted and observed
counting rates rather than source brightnesses from spectral fits,
particularly when energy resolution is limited, as with \rosat.
For an observation subtending a solid angle $\Delta\Omega$,
the counting rate is
$\Delta\Omega\:\sum_{il} B_{il}A_{il}$,
where $A_{il}$ is the instrument effective area at the 
energy of line $l$ and the sum is over all lines within
the chosen energy range.
For ACIS, one chip subtends 
$5.9 \times 10^{-6}$ steradian, or 69.8 arcmin$^{2}$,
and we find that the predicted ACIS-S3 rate between 511 and 886 eV, 
encompassing all the He-like and H-like O lines, is 
$9.0 \times 10^{-6}$ cts s$^{-1}$ arcmin$^{-2}$.

In comparison, the corresponding observed rate from the July S3
data, excluding ObsID 2469, 
is $34 \pm 20 \times 10^{-6}$ cts s$^{-1}$ arcmin$^{-2}$,
while the ObsID 2469 rate is 
$155 \pm 49 \times 10^{-6}$ cts s$^{-1}$ arcmin$^{-2}$.
The September rate, from the three brightest ObsIDs,
is $287 \pm 39 \times 10^{-6}$ cts s$^{-1}$ arcmin$^{-2}$,
a factor of 32 higher than predicted for an average solar wind,
in general agreement, as one would expect, with the results from 
the spectral fits in \S\ref{sec:modelspec} listed in
Table~\ref{table:fitbright}.

\subsection{Adjustments for Actual Solar Wind}
\label{sec:windadjust}

\begin{figure*}[h]
\centerline{\includegraphics[scale=.9,bb=38 66 566 754,clip]{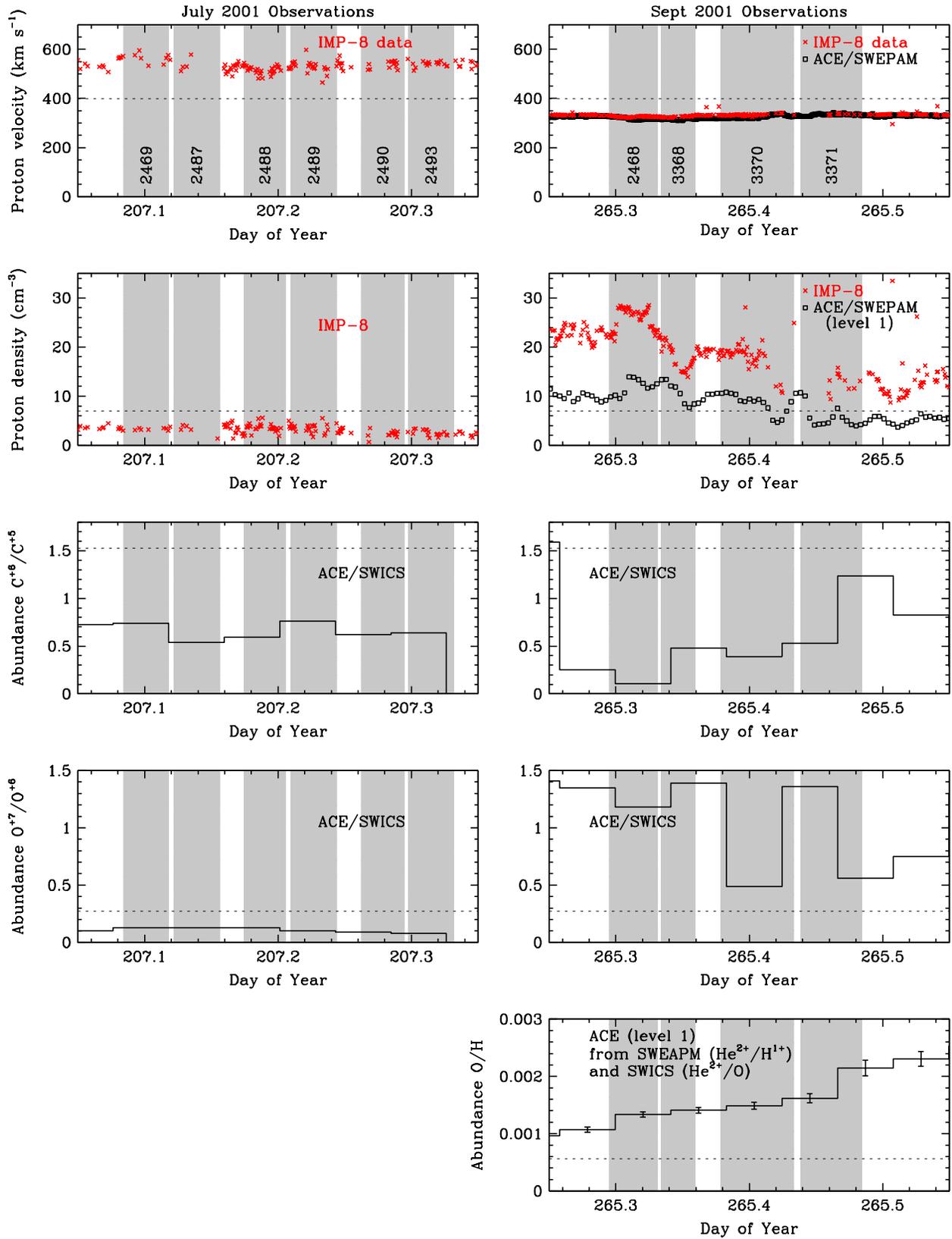}}

\caption{Solar wind data for \chandra\ observations.
Gray shading marks the duration of each ObsID.
Dotted horizontal lines mark average
values for the slow solar wind.
Level 1 data from \ace\ (in September proton density plot)
are not considered reliable, but are shown for
comparison with \imp\ data.
\imp\ and \ace\ proton velocity data in September are essentially
coincident.
Based on Eq.~1, and the corresponding data
plotted here, the September X-ray rates are predicted to
be much higher than in July, in agreement with observation.
}
\label{fig:winddata}
\end{figure*}

An obvious question raised by the preceding analysis is whether
or not the assumptions regarding solar wind parameters are appropriate
for the July and September observations.
This can be addressed by publically available data 
from solar monitoring instruments, specifically the 
{\it Interplanetary Monitoring Platform 8 (IMP-8)}\footnote{%
        Data available from 
        {ftp://space.mit.edu/pub/plasma/imp/www/imp.html}.},
and the Solar Wind Electon, Proton, and Alpha Monitor (SWEPAM)
and Solar Wind Ion Composition Spectrometer (SWICS) onboard the
{\it Advanced Composition Explorer (ACE)}\footnote{%
        Data available from {http://www.srl.caltech.edu/ACE/}.}.
Fig.~\ref{fig:winddata} shows the available relevant data,
namely the solar wind velocity and density, and various
element and ion relative abundances.
Horizontal dotted lines denote the average solar-wind parameter values
given previously.
It is immediately obvious that CT emission rates
should be much larger than average in September,
and smaller in July.

In July, the wind velocity $v_{c}$ averaged around 530 km s$^{-1}$,
versus 335 km s$^{-1}$ in September.  Wind densities, on the
other hand, are much higher in September than in July. 
The proton density measured by \imp\ in July 
is $\sim 3.5$ cm$^{-3}$, or half
the baseline value, while in September the density
ranges from $\sim26$ cm$^{-3}$ for the first ObsID (2468)
to $\sim13$ cm$^{-3}$ in the last (ObsID 3371).
The \ace/SWEPAM densities are about half as
large as the densities measured by \imp, but are only
available in September, and only for level-1 data which are
not recommended for scientific analyses.  \ace\ also orbits around the
Lagrange L1 point lying 0.01 AU ($\sim 240 R_{E}$) toward the Sun, 
whereas \imp\ is much closer to the Earth (orbital distance $\sim 35 R_{E}$).
For all those reasons, we therefore rely on the \imp\ densities.
Note that \ace\ data plotted in Fig.~\ref{fig:winddata}
have been time-shifted to account for wind travel-time to Earth:  
July data by $1.584 \times 10^{6}$ km $/$ 530 km s$^{-1} = 2990$ s,
and September data by
$1.435 \times 10^{6}$ km $/$ 335 km s$^{-1} = 4280$ s.
\imp\ data have at most a few hundred seconds of delay.
After these corrections, the time-behavior of the \imp\ and \ace\ 
density data match each other very well.

Oxygen abundance data are not available for July, but
the September data show a significant enhancement over
average values, by roughly a factor of two during ObsID 2468
and rising to more than a factor of three during ObsID 3371.
In July, \ace/SWICS data indicate a relatively lower ionization
state than usual for carbon (see Table~\ref{table:ions}).
This is also true of the September data, although the C$^{6+}$/C$^{5+}$
abundance ratio is more volatile.  There is a
significant rise in the 
ratio during ObsID 3371, 
which might be correlated with the increase in emission above 700 eV
associated with especially highly-charged ions during that observation
(see Table~\ref{table:ratestats}).
Overall, the abundance of the C$^{6+}$ ions that CT to
produce \ion{C}{6} Lyman emission lines is lower than usual,
in agreement with the relative weakness of those lines in
the September spectra.

\ace/SWICS does not measure the O$^{8+}$/O$^{7+}$ ratio,
but the O$^{7+}$/O$^{6+}$ ratio, which is usually 20:73 
(see Table~\ref{table:ions}),
is less than half that in July and much higher in September.
Determining what the absolute ion fractions are can not be
done precisely since the wind ions are not really in ionization equilibrium,
but based on the tabulations of \citet{cit:mazzotta1998} we estimate that
the O$^{6+}$:O$^{7+}$:O$^{8+}$ relative abundances are roughly
90:10:0 in July and 35:50:15 during most of the September
observations, with a drop in ionization level to 60:30:10
during most of ObsID 3370.  That drop in the abundance
of highly charged oxygen may be largely
responsible for the lower X-ray emission observed
during ObsID 3370 (see Table~\ref{table:ratestats}).

Putting all the above factors together ($v_{c}$, $n_{p}$,
$n_{O}/n_{p}$, and $f_{i}$)
and assuming that the July oxygen abundance was normal,
the CT X-ray intensity between 500 and 900 eV
in July should be 
0.25 times the baseline intensity, or
$2.3 \times 10^{-6}$ cts s$^{-1}$ arcmin$^{-2}$.
The predicted bright-September rate is 
15 times the baseline rate and 60
times the July rate, or
$140 \times 10^{-6}$ cts s$^{-1}$ arcmin$^{-2}$.
Considering the model uncertainties, particularly
our approximation for the neutral H density distribution,
this is very good agreement with the
observed quiet-July and bright-September rates of
$34 \pm 20$ and $287 \pm 39 \times 10^{-6}$ cts s$^{-1}$ arcmin$^{-2}$,
which are quoted with $1\sigma$ uncertainties.


\subsection{Application to \rosat\ Moon Observation}
\label{sec:rosatmoon}

We can also compare predicted and observed rates for the \rosat\
Moon data.
Because \rosat\ was in low-Earth orbit, observing perpendicular
to the Sun-Earth axis 
through the flanks of the magnetosphere,
$r_{min} = 15R_{E}$.
(Note that the line brightnesses plotted in Fig.~\ref{fig:modelspec}
are for this case, with the assumption of average solar wind parameters.)
\ace\ had not yet been launched, but during the
time of the observation, 1990 June 29  2:11--2:45 UT, 
\imp\ measured $v_{c}=400$ km s$^{-1}$
with a slightly elevated proton density of 8.5 cm$^{-3}$.
$B_{il}$ is then equal to 
$5.37 \times 10^{14} y_{il} f_{i} \sigma_{i}\;\;                
    \mathrm{phot}\; \mathrm{s}^{-1}\mathrm{cm}^{-2} \mathrm{sr}^{-1}$.

Using the \rosat\ PSPC effective area for $A_{il}$,
and with $\Delta\Omega = 3.0\times10^{-5}$ steradians for the half-Moon,
the observed rate, $\Delta\Omega\:\sum_{il} B_{il}A_{il}$,
is then predicted to be
0.0038 cts s$^{-1}$, vs the vignetting-corrected dark-side observed value 
of $\sim0.20$ cts s$^{-1}$ \citep{cit:schmitt1991}, 
a difference of roughly a factor of 50.
In the units of the \rosat\ All-Sky Survey,
the predicted CT rate is $11\times10^{-6}$ cts s$^{-1}$ arcmin$^{-2}$, while 
the observed dark-side rate is nearly 
$600\times10^{-6}$ cts s$^{-1}$ arcmin$^{-2}$.

Given this large discrepancy, 
we look more closely
at the \citet{cit:schmitt1991} observation.
In their Fig.~4 we see that
the intensity of full-band SXRB emission surrounding the Moon was 
$\sim 2000\times10^{-6}$ cts s$^{-1}$ arcmin$^{-2}$,
which is twice as high as the total rate of 
$\sim 1000\times10^{-6}$ cts s$^{-1}$ arcmin$^{-2}$ 
listed for the field of view in RASS
maps during the observation
(RA, Dec $\sim11^{h}44^{m}30^{s}$, $-03^{\circ}22'$).
\citet{cit:snowden1995} also noted this and concluded that
``the lunar observation occurred during the time of a strong LTE.''
Although \imp\ data on proton velocity and density do
not indicate anything out of the ordinary at that time, we have seen
from \ace\ data during the September 2001 \chandra\ observations
that other wind parameters such as oxygen abundance and
ionization level can have a very large effect on net
CT emission.
We therefore agree that the \rosat\ data indicate an LTE,
and deduce that the extra $\sim1000\times10^{-6}$ cts s$^{-1}$ arcmin$^{-2}$
in the measured SXRB rate comprises $\sim600\times10^{-6}$ 
from a large increase in the geocoronal CT rate and $\sim400\times10^{-6}$
from the associated transient excess of heliospheric CT emission.
The rough equivalence of the geocoronal
and heliospheric components of the LTE
is consistent with model predictions by \citet{cit:cravens2001}.

\section{IMPLICATIONS FOR THE SXRB}
\label{sec:sxrb}

\subsection{Heliospheric Charge Transfer}
\label{sec:helioCT}

Although geocoronal CT accounts for much if not most of the
X-ray emission during an LTE, and can be of roughly the same
intensity as the cosmic X-ray background,
its quiescent level is an order of magnitude or more
below that of the typical SXRB.
Heliospheric CT emission, however, is several times stronger
than quiescent geocoronal emission,
as was first shown by \citet{cit:cravens2000},
who presented a model for the
distribution of neutral H and He within the heliosphere.
Neutral gas is depleted near the Sun because of photoionization
and charge transfer with H and He in the solar wind, and densities
are higher upwind, with respect to the Sun's relative motion
through the Local Interstellar Cloud, than downwind.
Densities can be approximated by the relation
$n_{n}=n_{n0}e^{-\lambda/r}$, where
$n_{n0}$ is the asymptotic neutral density,
$r$ is distance from the Sun, and $\lambda$ is the depletion scale.

Outside the heliosphere, in the undisturbed interstellar medium,
the neutral H density is 0.20 cm$^{-3}$,
but behind the shock front the density drops by nearly a
factor of two, so that $n_{n0} = 0.12$ cm$^{-3}$
\citep{cit:gloeckler1997}.
Helium, in contrast, is largely unaffected by the shock,
and $n_{n0} = 0.015$ cm$^{-3}$.
Based on the theoretical work of \citet{cit:zank1999},
$\lambda$ for H is approximately 5 AU upwind, 
7 AU on the heliosphere flanks,
and very roughly 20 AU downwind. 
He is less depleted near the Sun and has less spatial variation, 
with $\lambda \sim$ 1 AU.
\citet{cit:robertson2003} have developed a more sophisticated
model for the distribution of neutral H and He in all directions, 
but our predictions of absolute CT emission should not be
significantly less accurate; in both cases, the estimated
uncertainty is roughly a factor of two or three.
Our calculations, however, keep track of each CT emission line
and the \rosat\ PSPC effective area for that line energy,
which will allow us to draw more specific conclusions
regarding the relative strength of emission in various energy bands.

For heliospheric CT emission, Eq.~3 thus becomes
\begin{equation}
B_{il} = \frac{1}{4\pi} n_{p} f_{O} f_{i} v_{c} y_{il} \sigma_{i}
        n_{n0} \int_{1\:\mathrm{AU}}^{\infty} 
        \left( \frac{1\;\mathrm{AU}}{r} \right)^{2} e^{-\lambda/r}\,dr,
\end{equation}
where we again approximate the look direction as radially outward from the Sun.
Proton density, $n_{p}$, 
decreases as the solar wind expands away from the
Sun, leading to the $r^{-2}$ factor in the integral.
If we isolate the ion-specific terms, $y_{il} f_{i} \sigma_{i}$
(and ignore changes in the
abundance of each ion species with distance from the Sun---ions
change charge with each CT collision,
but the path length for CT is many tens of AU),
and evaluate the rest of Eq.~7 using the average solar wind
and neutral gas parameters listed previously,
we obtain
\begin{equation}
B_{il} = C y_{il} f_{i} \sigma_{i}\;\;
        \mathrm{phot}\; \mathrm{s}^{-1}\mathrm{cm}^{-2} \mathrm{sr}^{-1},
\end{equation}
where $C$ has units of s$^{-1}$ cm$^{-4}$ sr$^{-1}$ and parametrizes
the solar wind density and velocity and the neutral gas distribution
along the line of sight.
For CT with He, $C = 1.76 \times 10^{15}$, 
and for H it equals
$4.37 \times 10^{15}$, $3.12 \times 10^{15}$, and $1.06 \times 10^{15}$
when looking upwind, on the flanks, and downwind, respectively.
Observations that look through the
helium focusing cone downwind of the Sun, where He
density is enhanced by roughly a factor of four \citep{cit:michaels2002}, 
will see more
emission, with $C \sim 10^{16}$.  Note, however, that cross sections
for He CT with the most important ions are roughly a factor of two
smaller than for H CT (see Table~\ref{table:ions}, which
also lists values of $f_{i}$ for the slow wind).

\tabletypesize{\small}
\begin{deluxetable}{llc}
\tablecaption{Model CT Rate Predictions\label{table:snowdens}}
\tablewidth{0.9\linewidth}
\tablehead{
\colhead{Neutral}       & \colhead{Look}        & \colhead{Total \rosat\ Rate} \\
\colhead{Gas}           & \colhead{Direction} 
                & \colhead{($10^{-6}$ cts s$^{-1}$ arcmin$^{-2}$)}
}
\startdata
~~Helio H & upwind        & 87    \\
~~Helio H & flanks        & 62    \\
~~Helio H & downwind      & 21    \\
~~Helio He & any          & 16    \\
~~Helio He & He cone      & $\sim90$      \\
~~Geo H   & upwind        & 24    \\
~~Geo H   & flanks        & 11    \\
~~Geo H   & downwind      & $\sim0$
\enddata
\tablecomments{Results are for look directions within the slow solar wind.
\rosat\ always looked through the flanks of the Earth's magnetosphere;
geocoronal rates for upwind and downwind directions 
are provided for comparison and application to other missions.
Heliospheric emission when looking primarily through the fast wind is 
much lower.}
\end{deluxetable}

Predicted \rosat\ counting rates from quiescent CT emission
are listed in Table~\ref{table:snowdens}.
The total of geocoronal and heliospheric emission is around
$80 \times10^{-6}$ cts s$^{-1}$ arcmin$^{-2}$, varying by
 $\pm30 \times10^{-6}$ cts s$^{-1}$ arcmin$^{-2}$
depending on look direction through the
slow solar wind.  The highest rate, roughly 
$120 \times10^{-6}$ cts s$^{-1}$ arcmin$^{-2}$,
will be observed
when looking through the He cone.
For observations primarily through the fast wind, the
geocoronal component remains the same, while the He and
especially H heliospheric components will be significantly reduced,
yielding a total of
$\sim 25 \times10^{-6}$ cts s$^{-1}$ arcmin$^{-2}$.
For comparison, 
typical RASS rates away from bright Galactic structures
are 600, 400, and 100 $\times 10^{-6}$ cts s$^{-1}$ arcmin$^{-2}$
for the full, R12, and R45 energy bands, respectively.
For lines of sight mostly through the slow wind
(which is most of the sky during solar maximum, when
the \rosat\ background maps were conducted),
our model therefore predicts that
combined quiescent geocoronal and heliospheric CX emission 
accounts for roughly 13\% of the total 
SXRB measured by \rosat, 11\% of the R12-band emission, and
one-third of the R45-band emission.
Overall uncertainty in our predictions is roughly
a factor of two.

Slightly more than half of the total CT emission ($\sim55$\%)
is predicted to be in the R12 band,
which is consistent with the observation that 
the \rosat\ LTEs appear most strongly in the R12 band 
\citep{cit:snowden1995}.
The true R12-band emission is almost certainly even higher, however, 
because of the known incompleteness of our
CT spectral model.
Although we model Li-like \ion{O}{6} and \ion{Ne}{8} emission, 
there is additional
$L$-shell ($n \rightarrow 2$) CT emission, mostly in the
R12 band, resulting from 
Li-like and lower charge states of Mg, Si, S, Ar, and Fe.
Although more difficult to model,
these heavier elements have significant abundance
(in sum, approximately the same as for 
bare and H-like C \citep{cit:schwadron2000}),
and they have been invoked to explain
much of the low-energy emission seen in comets 
\citep{cit:krasno2004}.

Also note that some of the SXRB recorded in the RASS comes from
point sources which could not be resolved
by \rosat\---\citet{cit:maxim2002} estimate 20\%---so
the fraction of the ``true'', i.e., diffuse, SXRB arising from CT is
correspondingly higher.
Whatever the average level of CT emission is,
we expect that it may well be
the dominant contributor to the SXRB in some low-intensity fields,
particularly in the R45 (3/4-keV) band.
If so, CT emission will have a significant impact on
our understanding of the SXRB and models of the local interstellar medium,
particularly the Local Bubble.

\subsection{Once and Future Data}

\rosat, of course, is not the only X-ray satellite affected
by CT emission; {\em any} X-ray observation made from
within the solar system will be impacted.
Variations in the strength of background O emission
have been noted (tentatively) in repeated \chandra\ observations of
MBM~12, a nearby dark molecular cloud that shadows much
of the SXRB (R.~J.\ Edgar et al., in preparation), 
and in {\it XMM-Newton} observations
of the Hubble Deep Field North, which also showed variable Ne 
and Mg emission
(S.~L.\ Snowden, 2004, private communication).
Such variations may be important when observing
weak extended sources such as clusters
where background determinations are critical.

With future missions
having much better energy resolution, such as {\it ASTRO-E2}
with its 6-eV-FWHM-resolution microcalorimeter array 
\citep{cit:stahle2003},
one should be able to see differences in 
background spectra between areas of the sky corresponding to
low and high solar latitudes (around solar minimum)
because of the difference in ionization levels
between the slow and fast solar winds.
It should also be easy to identify spectral features indicative
of CT, such as an anomalously strong \ioka\ forbidden line
($1s2s$ $^{3}S_{1} \rightarrow 1s^{2}$ $^{1}S_{0}$).
When excited by CT, that line is
several times stronger than the
resonance line ($1s2p$ $^{1}P_{1} \rightarrow 1s^{2}$ $^{1}S_{0}$),
completely unlike their ratio in thermal plasmas
\citep{cit:khar2003}.
Other signatures of CT are
the high-$n$ Lyman transitions of O and C,
which are strongly enhanced relative to Ly$\alpha$.
Indeed, there are hints of high-$n$ O Lyman lines in a short (100 sec)
microcalorimeter observation \citep{cit:mccammon2002}.
There is also a strong feature at $\sim67.4$ \AA\ (184 eV) in the
{\it Diffuse X-Ray Spectrometer} ({\it DXS}) spectrum of the SXRB
\citep{cit:sanders2001}
which cannot be reproduced by thermal plasma models.
A promising explanation is CT emission at 65.89 and 67.79 \AA\
(188.16 and 182.89 eV) from 
Li-like \ion{Ne}{8} $1s^{2}5d \rightarrow 1s^{2}2p$ 
and H-like \ion{O}{8} $5d,5s \rightarrow 2p$ transitions, which are the
strongest CT features 
within the 148-284-eV DXS energy range (see Fig.~\ref{fig:modelspec}, and
note that the plotted resolution is very similar to that for 
{\it ASTRO-E2} and {\it DXS}).

Although variability in foreground emission, 
whether from changes in the solar wind or
in the position of the observing spacecraft with respect to
the magnetosphere and heliosphere,
can be an annoying complication when analyzing extended sources,
variations of CT emission in time and space
also provide opportunities to learn more about the solar wind,
geocorona, and heliosphere.  Data from \chandra\
and {\it XMM-Newton} will provide important constraints
on our models of CT emission, and future missions
such as {\it ASTRO-E2} should permit definitive tests.
Given the unavoidable and rather large uncertainties in 
current theoretical models,
observational data will be critical in this endeavor.

\section{CONCLUSIONS}
\label{sec:conclusions}

As described in this paper,
we have detected significant time-variable soft X-ray emission
in \chandra\ observations of the dark side of the Moon 
which is well explained by our model of geocoronal
charge transfer.
The observed brightness ranged from a maximum of
$\sim 2 \times 10^{-6}$ phot s$^{-1}$ arcmin$^{-2}$ cm$^{-2}$,
with most of the emission 
between 500 and 900 eV, to a minimum at least an order of magnitude lower.
Predicted intensities, which are based in part on detailed solar wind data,
match observation
to within a factor of two, which is within the model uncertainty.
Emission from \ioka, \iolya, and a blend of high-$n$ \ion{O}{8} Lyman lines 
is detected with high confidence, as well as probably \imgka\ and
perhaps high-$n$ emission from \ion{C}{6}.
We also include estimates of heliospheric emission
and find that the total charge transfer emission amounts to a
substantial fraction of the soft X-ray background, roughly
one-third of the rate measured in the \rosat\ R45 band.
Future observations with microcalorimeter detectors should
allow much more accurate assessments of the contribution of
charge transfer emission to the SXRB because of its unique spectral signatures.

 
\acknowledgments

We gratefully acknowledge very helpful discussions
with C.~Grant, P.~Plucinsky, J.~Raymond, and S.~Snowden.
We also thank the \ace\ SWEPAM and SWICS instrument teams 
and the \ace\ Science Center for providing their data, 
and the MIT Space Plasma Physics Group
for the \imp\ data.
The authors were supported by 
NASA's {\it Chandra X-ray Center} Archival Research Program 
under Grant AR4-5001X during the course of this research.  
BW, MM, MJ, and RE were also supported by 
NASA contract NAS8-39073 to the {\it CXC},
and AD and VK were supported by the NASA Planetary
Science Program under Grant NAG5-13331.





\end{document}